\documentclass[aps, pra, twocolumn, superscriptaddress, amsmath, tightenlines, longbibliography]{revtex4-2}

\newcommand{\RNum}[1]{\uppercase\expandafter{\romannumeral #1\relax}}
\usepackage{color}

\usepackage{amssymb}
\usepackage{amsmath}
\usepackage{dcolumn}
\usepackage{graphicx}
\usepackage{mathrsfs}
\usepackage{appendix}
\usepackage{graphicx}
\usepackage{booktabs}
\usepackage{colortbl}

\definecolor{Dred}{RGB}{190,0,0}

\usepackage{url}
\usepackage[colorlinks]{hyperref}
\hypersetup{%
	plainpages=true,
	breaklinks=true,       
	hypertexnames=false,  
	pageanchor=true,
	colorlinks=true,
	linkcolor={blue},
	citecolor={red},
	urlcolor={blue},
	anchorcolor={black}
}

\def \hide#1{}

\begin{document}

\title{Light-matter interactions in a Hofstadter lattice with next-nearest-neighbor couplings}

\author{Jia-Qi Li}
\affiliation{Institute of Theoretical Physics, School of Physics, Xi'an Jiaotong University, Xi'an 710049, People’s Republic of China}

\author{Zhao-Min Gao}
\affiliation{Institute of Theoretical Physics, School of Physics, Xi'an Jiaotong University, Xi'an 710049, People’s Republic of China}

\author{Wen-Xiao Liu}
\affiliation{Department of Electronic Engineering, North China University of Water Resources and Electric Power, Zhengzhou 450046, People’s Republic of China}
\affiliation{Institute of Theoretical Physics, School of Physics, Xi'an Jiaotong University, Xi'an 710049, People’s Republic of China}

\author{Xin Wang}\email{wangxin.phy@xjtu.edu.cn}
\affiliation{Institute of Theoretical Physics, School of Physics, Xi'an Jiaotong University, Xi'an 710049, People’s Republic of China}

\date{\today}

\begin{abstract}
The light-mater interactions for an emitter coupling to the bulk region of a Hofstadter lattice have been recently investigated by De Bernardis \textit{et al.} [D. De Bernardis, Z.-P. Cian, I. Carusotto, M. Hafezi, and P. Rabl, \href{https://link.aps.org/doi/10.1103/PhysRevLett.126.103603}{Phys. Rev. Lett. 126, 103603 (2021)}]. We propose the light-mater interactions in an extended Hofstadter lattice with the next-nearest neighbor (NNN) couplings. Compared with the standard Hofstadter lattice, the NNN couplings break the mirror symmetry and the energy bands are not flat, i.e., dispersive with nonzero group velocity. 
In contrast to the study by De Bernardis \textit{et al.}, when a two-level emitter interacts 
with the bulk region of the extended Hofstadter lattice, the emitter is no longer trapped by the  
coherent oscillations, and can radiate photons unidirectional. The chiral 
mechanism stems from the broken mirror symmetry. Both the radiation rate and the chirality 
periodically change with the emitter's coupling position. All of those particular features can 
be realized on the photonic lattice platform and may find potential application in chiral 
quantum information processing.
\end{abstract}

\maketitle

\section{introduction}
Exploring the interaction between quantum emitters and photonic baths with various kinds of spectra is the central topic of quantum optics~\cite{Claude1992}. One well-known phenomenon is spontaneous emission, i.e., an excited emitter spontaneously emits the energy into the environment. By designing the bath with finite spectral bandwidth, a large amount of interesting phenomena beyond spontaneous emission in quantum electrodynamics (QED) are observed. For example, non-Markovian evolution~\cite{Bykov1975,Yablonovitch1987,Hoeppe2012,GonzlezTudela2018,Stewart2020,Ferreira2021} and bound states~\cite{John1990,Lambropoulos2000,Plotnik2011,Douglas2015,Krinner2018,Vega2021,Scigliuzzo2022} are demonstrated in the photonic crystal waveguide with a band gap~\cite{Joannopoulos2011,Chen2014}. Moreover, when the bath is spatiotemporally modulated, analogue Hawking radiation~\cite{Nation2009,Tian2019,Katayama2020} and chiral transport~\cite{Ramos2016,Barik2018,Wang2022,Siampour2023,Barik2020} have been realized. All those important processes indicate that the specific artificial platform, where the structures of photonic bath are designed, can be used to demonstrate the quantum optics.

Recently, unconventional quantum phenomena with the lattice model in condensed matter physics, 
have attracted great interest~\cite{Bello2019,Ozawa2019,GarciaElcano2020,Poshakinskiy2021,Rivera2020,Leonforte2021,
Ruks2022,Miguel2023,Fernandez2022,Cheng2022,Du2022}. Those lattices usually have nontrivial spectra 
and extraordinary topological properties. For example, when an emitter resonates with the central frequency of a two-dimensional square lattice's band, the exponential decay rate does not obey the Fermi's golden rule, but is predicted by overdamped oscillations and slow relaxation dynamics~\cite{Cirac2017PRA,Cirac2017PRL}. Moreover, considering a small emitter coupling to a hexagonal lattice and tuning in the Dirac point, the decay follows a logarithmic law~\cite{Gonzalez2018}. All those phenomena show that the lattices with nontrivial spectra provide versatile platforms for exploring QED phenomena beyond conventional photonic baths.

In condensed matter, when the lattice lies in a magnetic field, the spectrum becomes non-trivial, leading to the quantum/fractional Hall effect~\cite{Klitzing1980,Tsui1982,Laughlin1983,Halperin1982,vonKlitzing2020} and topologically protected edge states~\cite{Hafezi2011,Hafezi2013,Poo2011,Fang2012,Khanikaev2017,Wang2017,Liu2019}. For a square lattice with a perpendicular gauge field, an elegant fractal structure spectrum known as the Hofstadter butterfly emerges~\cite{Hofstadter1976}. In Ref.~\cite{Bernardis2021}, by considering an emitter coupling to the bulk region of a Hofstadter model (H-model), the emission process displays coherent oscillations or no decay at all. When an emitter is located at the edge of the H-model and resonant with different band-gaps, the emission characteristics become quasi-quantized and chiral due to the increasing edge modes~\cite{Vega2022}. The magnetic field not only leads to so-called Landau levels, but also causes the topologically protected chiral edge states~\cite{Landau1981,Harper1955,Raghu2008,Haldane2008}. 

\begin{figure*}[tbph]
	\centering \includegraphics[width=16.8cm]{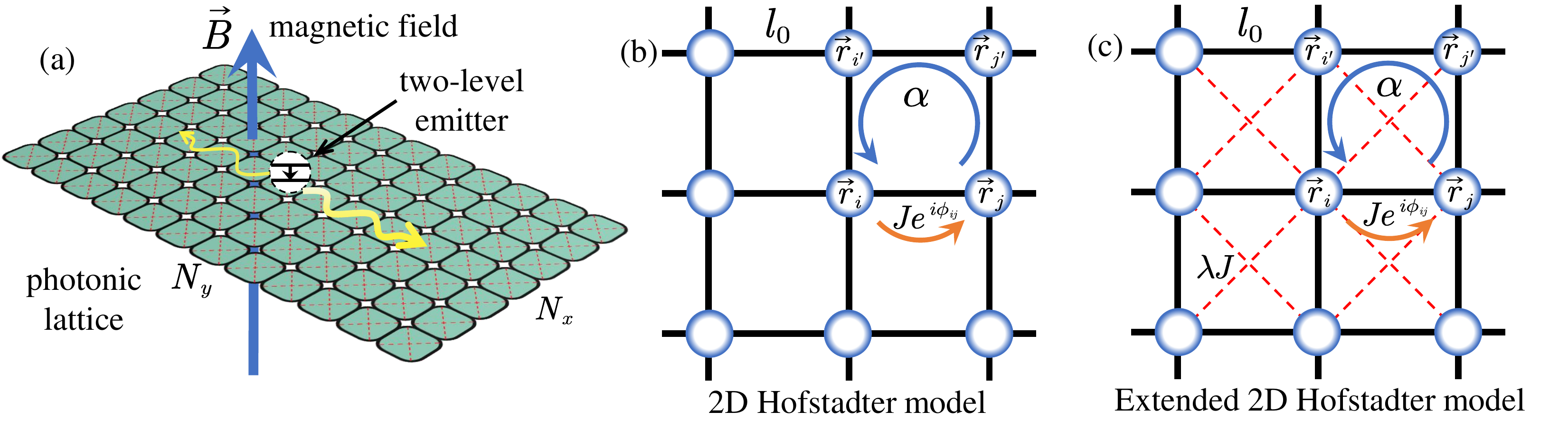}
	\caption{(a) Sketch of a setup of a two-level emitter coupling to the bulk region of a photonic lattice, which  corresponds to an extended Hofstadter model with the next-nearest-neighbor (NNN) hopping (red dashed lines), under a perpendicular synthetic magnetic field $\vec{B}$. (b) Schematic of the 2D Hofstadter model (H-model), i.e., a square photonic lattice with hopping strength $J$ and hopping phase $\phi_{i,j}$ between nerighboring lattice sites. For each plaquette, $\sum{_{\Box}\phi _{i,j}=2\pi \alpha}$. (c) The extended 2D H-model with the NNN couplings $\lambda J$ between diagonal sites of the square.
	}
	\label{fig1}
\end{figure*}

In experiments, the H-model is often realized in an artificial platform with tunable 
nearest-neighbor (NN) hopping~\cite{Hafezi2011,Hafezi2013,Jaksch2003,Aidelsburger2013,Miyake2013,Dean2013,Aidelsburger2014,Roushan2017}. However, when the lattice sites become closer to each other,   the parasitic next-nearest neighbor 
(NNN) couplings must be considered~\cite{Yan2018,ZhaoPeng2020}. In the zigzag waveguide arrays, when the distance between 
the waveguides is twice of that between the layers, the NNN couplings strength is $30\%$ of 
the NN coupling~\cite{Dreisow2008}. In this scenario, the NNN couplings take significant 
effects and lead to new physical phenomena beyond the NN approximation~\cite{Wang2010,Dreisow2011,Sasaki2006,Bellec2013,Chakraborty2010,Chandler2016,Schulz2022,Du2021,SerraGarcia2018}.

In this paper, we consider the extended H-model with the NNN couplings as a perturbation and 
investigate the quantum dynamics of an emitter coupling to the bulk region of the lattice. The NNN sites are coupled via a real tunneling amplitude, without hopping phases. The mirror symmetry is broken by the NNN couplings both in the $x$ and $y$ directions (see Appendix A)~\cite{Benalcazar2017,SerraGarcia2018,Schulz2022}. Consequently, the flat bands become dispersive with nonzero group 
velocity. When the emitter resonates with the middle of the lowest band, the decay follows an exponential law. More intriguingly, the emission field shows periodicity and strong chirality, which paves the way for realizing chiral quantum optics~\cite{Petersen2014,Lodahl2017,Calaj2019,Wang2022b,Wang2022njp}. Different from the widely discussed chiral edge state in the H-model~\cite{Raghu2008,Wang2009,Kudyshev2019,Vega2022}, our findings about the chiral emission in the bulk region of the lattice are rarely studied~\cite{Bernardis2023}. Our proposal indicates that the unavoidable NNN couplings play an important role in an artificial photonic lattice, 
and can lead to unconventional QED phenomena beyond standard platforms.

The structure of this paper is as follows: in Sec.~\RNum{2}, we introduce the extended H-model with the NNN couplings, and obtain dispersion relations via the quasi-continuous Harper equation.
In Sec.~\RNum{3}, we consider an emitter coupling to the extended H-model, and derive the spontaneous emission rate under the Markovian approximation. In Sec.~\RNum{4}, by analyzing distribution properties of the lattice modes, we explain the mechanism of chiral emission and its periodicity. In Sec.~\RNum{5}, we summarize our main results.   

\section{Model}

We consider a two-level emitter with frequency $w_e$ located in the bulk region of an extended Hofstadter model, which contains additional NNN hopping between the diagonal sites of the lattice, as shown in Fig.~\ref{fig1}(a). The lattice constant is set as $l_0=1$ and $N_x \times N_y$ is the total number of lattice sites. Each lattice site is assumed as a photonic cavity. The annihilation operator at the site $\vec{r}=\left(x,y\right)$ is denoted as $a_{x,y}$. The coupling strength between the  nearest-neighboring sites is $Je^{i\phi_{ij}}$. The NNN hopping rate is $\lambda J$ and assumed to be identical and real. Therefore, the Hamiltonian of the extended H-model is written as ($\hbar =1$ and $J=1$)
\begin{eqnarray}
H_m=&-&\sum_{x,y} \left[\left( a_{x+1,y}^{\dagger}a_{x,y}+e^{i\phi _{x,y}}a_{x,y+1}^{\dagger}a_{x,y} \right) \right.\notag \\
&+&\left. \lambda\left( a_{x+1,y+1}^{\dagger}a_{x,y}+a_{x+1,y-1}^{\dagger}a_{x,y} \right) \right] +\mathrm{H}.\mathrm{c}.
\label{H_tight}
\end{eqnarray}
Here the phase $\phi _{i,j}=\frac{e}{\hbar}\int_{\vec{r}_j}^{\vec{r}_i}{\vec{A}\left( \vec{r} \right)}\cdot d\vec{r}$ originates from a synthetic magnetic field $\vec{B}=\nabla \times \vec{A}$, where we employ the Landau gauge with $\vec{A}=B(0,x,0)$. We define a dimensionless parameter $\alpha$ to denote the effective magnetic flux~\cite{Bernardis2021,Fang2012,Vega2022}
$$
\alpha = \frac{1}{2\pi}\sum{_{\Box}\phi _{ij}}=\frac{1}{2\pi}(\phi_{i,j}+\phi_{j,j'}+\phi_{j',i'}+\phi_{i',i})=\frac{e\Phi}{2\pi \hbar},
$$
where $\sum_{\Box}$ represents the summation of the hopping phase in a unit cell [i.e., the blue arrow in Fig.~\ref{fig1}(b)]. In this case, $\Phi =Bl_{0}^{2}$ is the flux enclosed in per plaquette. Without loss of generality, we take $\alpha=1/M$ with $M\in \mathbb{N}$. 

In the presence of Landau gauge, the spectrum of the extended H-model in the $k_y$ direction is not flat with nonzero group velocity. In the $k_x$ direction, the group velocity is still zero. We adopt a perturbation approach to obtain the analytical dispersion relation. We assume that the wavefunction can be written as $\psi (x,y) =e^{ik_yy}u(x)$, with pure Bloch waves in y-axis. By substituting $\psi (x,y)$ into Schrödinger equation $H_m\psi(x,y)=E\psi(x,y)$, the discrete extended Harper equation is derived as~\cite{Harper1955,Harper2014}
\begin{eqnarray}
E u(x)=&-&2\cos \left( 2\pi \alpha x-k_y \right) u(x)\notag \\
&-&(1+2\lambda \cos k_y) \left[u(x+1)+ u(x-1)\right],
\label{Harper_equation}
\end{eqnarray}
which indicates that $u(x)$ is periodic with $1/\alpha=M$. Therefore, we truncate this lattice to $x=M$, and apply periodic boundary conditions. 

\begin{figure}[tbph]
	\centering \includegraphics[width=8.6cm]{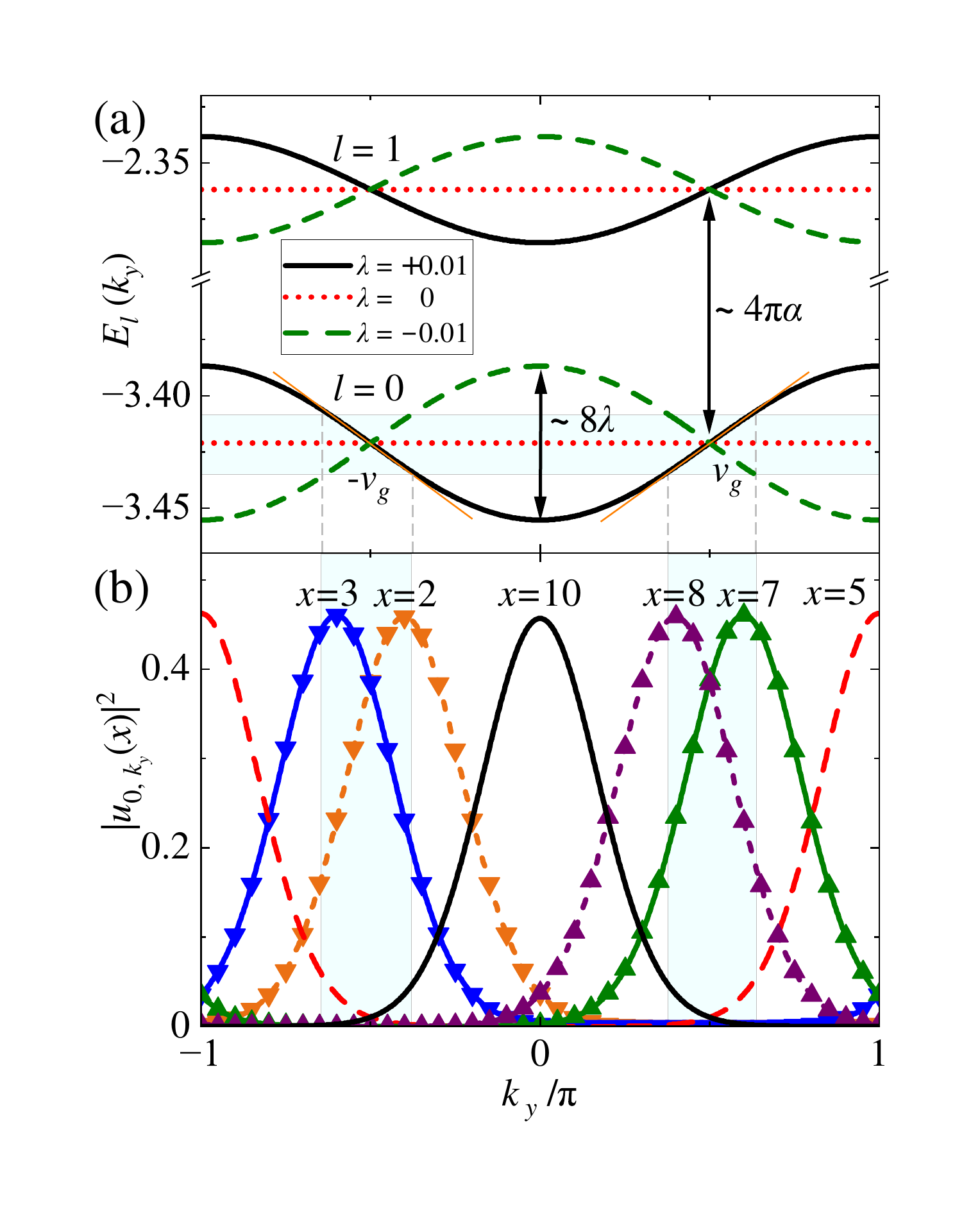}
	\caption{(a) The dispersion relations for the two lowest energy levels with $\lambda=0$, and $\pm 0.01$. We assume that the emitter's frequency lies in the cyan area. In the middle of the lowest band, the group velocity is $\pm v_g$. (b) The probability $|u_{l,k_y}(x)|^2$ versus $k_y$ for $l=0$ and $\lambda=0.01$ with different $x$ values. The parameter for those plots is $\alpha=1/10$.
	}
	\label{fig2}
\end{figure}

We adopt the quasi-continuum approximation to replace $u(x\pm 1)$ with $e^{\pm \partial_{x}}u(x)$ (see more details in Appendix B)~\cite{Harper2014}. Consequently, the dispersion relations are simplified as: 
\begin{widetext}
\begin{gather}
E_{l,k_y} \simeq -4-4\lambda \cos k_y
+\left[ 2\pi\alpha\left( l+\frac{1}{2} \right) -\frac{1}{16}\left( 2\pi \alpha  \right) ^2\left( 2l^2+2l+1 \right) \right] 2\sqrt{1+2\lambda \cos k_y}, \label{dispersing_relation1}
\end{gather}
\end{widetext}
where $E_{l}$ is the $l$th Landau level. For higher energy levels, this perturbation approach is invalid. If $\lambda=0$, the cosine terms in Eq.~(\ref{dispersing_relation1}) vanish and the eigenvalues become constants with high degeneracy, i.e., $$E_{l,k_y} \simeq -4
+2\left[ 2\pi\alpha\left( l+\frac{1}{2} \right) -\frac{1}{16}\left( 2\pi \alpha  \right) ^2\left( 2l^2+2l+1 \right) \right].$$ However, when considering nonzero NNN couplings, the eigenvalues vary with $k_y$. The energy levels become dispersive bands with a bandwidth $$W_l=E_{l,k_y=\pi}-E_{l,k_y=0}\simeq 8\lambda - 8\pi \alpha \left( l+\frac{1}{2} \right).$$ In Fig.~\ref{fig2}(a), the red horizontal curves ($\lambda=0$) and the black curves ($\lambda=0.01$) are the flat bands of the H-model and the dispersive bands of the extended H-model, respectively. The NNN couplings break the mirror-symmetry, which leads to the Landau levels nondegenerate and each band non-flat. Furthermore, The group velocity $v_g$ is nonzero, indicating that the photonic current can propagate along y-axis. 

The Hamiltonian is rewritten as a $M \times M$ matrix (see Appendix C). After numerical diagonalization, the $l$th energy band's discrete wavefunction $u_{l,k_y}$ is expressed as a column vector $$E_l: u_{l,k_y}=[u_{l,k_y}(1),u_{l,k_y}(2),...,u_{l,k_y}(M)]^{T}.$$ We plot the probability $|u_{l,k_y} (x)|^2$ for $l=0$ and $\alpha=1/M=1/10$ in Fig.~\ref{fig2}(b). The probabilities are asymmetric about $k_y=0$ [$u_{l,k_y}(x) \ne u_{l,-k_y}(x)$], except for $x=5$ and $10$. Note that for $\lambda=0$ the probability $|u_{l,k_y}(x)|^2$ still keeps this form. When the emitter couples to those points with asymmetric wavefunction, the emission is unidirectional. For simplicity, we replace $k_y$ with $k$ in the following discussion.

\section{SPONTANEOUS EMISSION}
\begin{figure}[tbph]
	\centering \includegraphics[width=8.6cm]{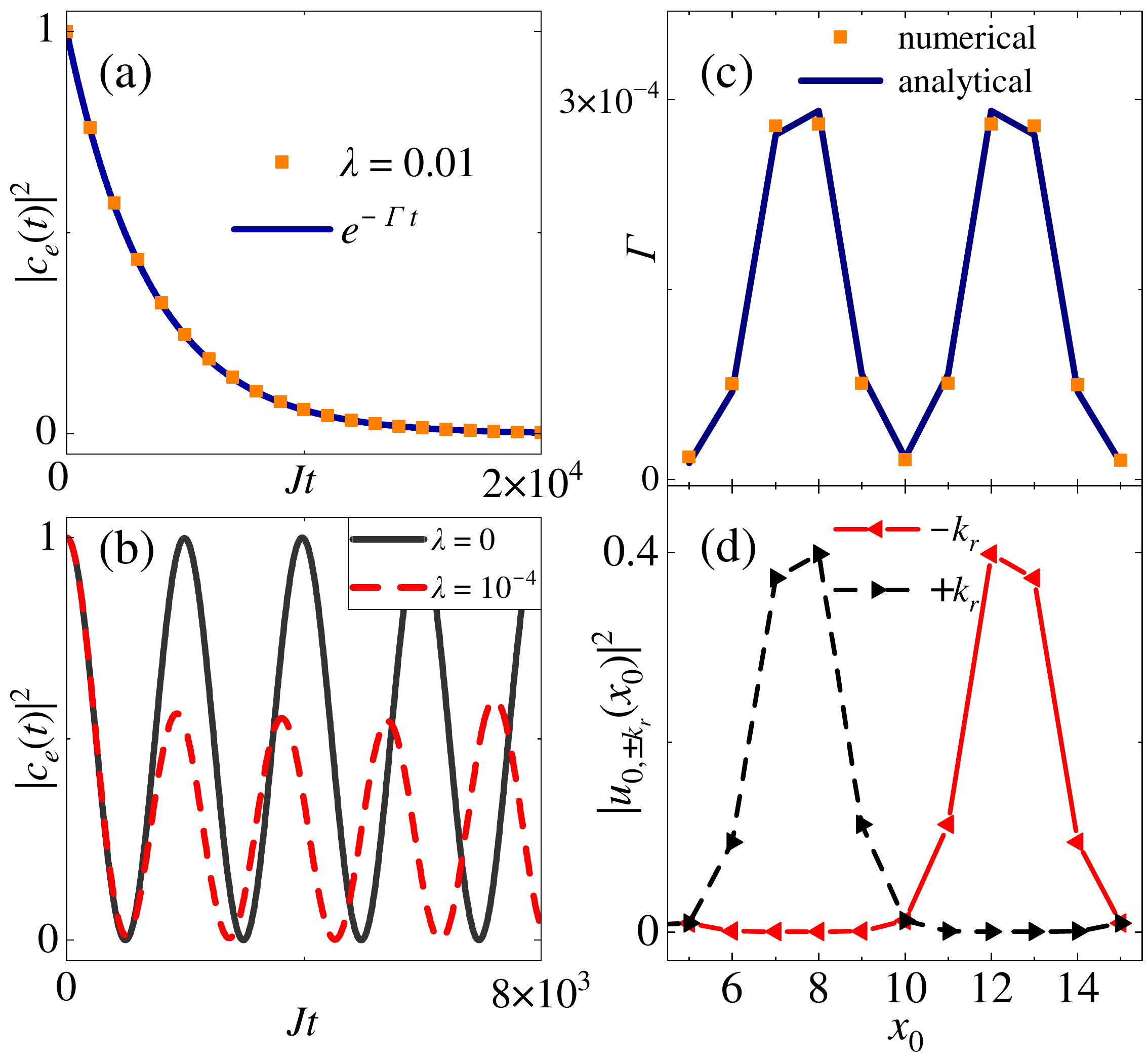}
	\caption{The probability $|c_e(t)|^2$ for (a) Markovian and (b) non-Markovian situation with different $\lambda$. The exponential decay rate $\varGamma $ is derived from Eq.~(\ref{rate}). The orange squares in (a) and two curves in (b) are all numerical calculations. The emitter is located at the position $(x_0=12, y_0=0)$ for a $25 \times 1000$ lattice. (c) The decay rate $\varGamma$ versus the coupling position $x_0$ with $\lambda=0.01$. The orange square dots and solid curve correspond to the numerical and analytical results, respectively. (d) The probability $|u_{0,\pm k_r}(x_0)|^2$ versus $x_0$ with $k_r=\pi/2$. The parameters for those plots are $\omega_q=-3.42$, $\alpha=0.1$ and $g=0.005$.}
	\label{fig3}
\end{figure}

We consider a two-level emitter located in the bulk region $(x_0,y_0)$ of the extended H-model [see Fig.~\ref{fig1}(a)]. The system's Hamiltonian is written as 
\begin{gather}
H_S = H_0+H_{\mathrm{int}},\\
H_0 = \frac{1}{2} w_q \sigma_z+H_m,\quad 
H_{\mathrm{int}}=g\left( \sigma _-a_{x_0 ,y_0}^{\dagger}+\sigma _+a_{x_0 ,y_0}\right), \label{H_int}
\end{gather}
where $w_q$ is the frequency of the emitter, $g$ is the interaction strength between the emitter and the lattice, and $\sigma _{z/\pm}$ are the Pauli operators of the atom. As discussed in Appendix C, the real space operator $a_{x_0,y_0}^{\dagger}$ is expressed as the combination of eigenmodes $X_{i,k}$. Consequently, the interaction Hamiltonian is rewritten as: 
\begin{gather}
H_{\mathrm{int}}=\frac{g}{\sqrt{N}}\sigma _-\sum_k\sum_{i=0}^M e^{-iky_0}u_{i,k}\left( x_0 \right) X_{i,k}^{\dagger} +\mathrm{H}.\mathrm{c}. \label{firstinter}
\end{gather}
We assume that the emitter resonates with the lowest band $E_{0,k}$. The band gap ($\simeq 4\pi\alpha$) between the two lowest bands is much larger than the bandwidth ($\simeq8\lambda$) of $E_{0,k}$~\cite{Bernardis2021}. The effects of the higher bands $E_{l}, l>0$ on the dynamics can be ignored. Therefore, Eq.~(\ref{firstinter}) is simplified as 
\begin{gather}
H_{\mathrm{int}}=\frac{g}{\sqrt{N}}\sum_k{e^{-iky_0}\sigma _- u_{0,k}(x_0)X_{0,k}^{\dagger}+\mathrm{H}.\mathrm{c}.}
\label{sim_H}
\end{gather}
Then, by replacing the eigenmode $X_{0,k}^{\dagger}$ with the creation operator in $k$ space (see Appendix C), we obtain
\begin{gather}
H_{\mathrm{int}}=\frac{g}{\sqrt{N}}\sum_{x =1}^M \sum_k e^{-iky_0}u_{0,k}(x_0)u_{0,k}(x)\sigma _-a_{x ,k}^{\dagger}+\mathrm{H}.\mathrm{c}.
\label{interactionH}
\end{gather}
There are two emission channels by assuming a resonant position at $w_q=E_{0,\pm k_r}$.  To continue, we apply the unitary transformation $U_0(t)=e^{-i H_0 t}$. The interaction operator becomes $$\sum_k {\sigma_- a_{x,k}^{\dagger}} \rightarrow \frac{N}{2\pi}\int_{-\pi}^{\pi}{\left( \sigma _-a_{x,k}^{\dagger}e^{i\Delta_k t} \right) dk},$$ where $\Delta_k=E_{0,k}-w_q$. 

We assume that initially the emitter is in the excited state, and the lattice modes are in the vacuum state. Therefore, The system's state in momentum space is expressed as
\begin{gather}
|\Psi (t) \rangle =c_e\left( t \right) |e,0\rangle +\sum_{x=1}^M \sum_k c_{x,k}\left( t \right)|g,1_{x,k}\rangle.
\label{statevector}
\end{gather}
Note that $c_{x,k}(t)$ is the probability amplitude of the photon located at $x$ with mode $k$. Substituting the state vector $|\Psi(t) \rangle$ into the Schrödinger equation, we derive
\begin{gather}
\dot{c}_e\left( t \right) =-i\frac{g}{\sqrt{N}}\sum_{x =1}^M\sum_ke^{iky_0}u_{0,k}(x_0)u_{0,k}(x)c_{x,k}\left( t \right)e^{-i\Delta _kt}, \label{dot(c_e)}
\\
\dot{c}_{x,k}\left( t \right) =-i\frac{g}{\sqrt{N}}e^{-iky_0}u^{*}_{0,k}(x_0)u^{*}_{0,k}(x)c_e\left( t \right) e^{i\Delta _kt} \label{dot(c_g)}.
\end{gather}
We first integrate Eq.~(\ref{dot(c_g)}) and substitute the integrated form into Eq.~(\ref{dot(c_e)}). The evolution equation of $c_e(t)$ is simplified as
\begin{gather}
\dot{c}_e\left( t \right) =-\frac{g^2}{2\pi}\int_{-\pi}^{\pi}{dk|u_{0,k}(x_0) |^2}\int_0^t{c_e\left( t' \right) e^{-i\Delta _k(t-t')}dt'},
\end{gather}
where we utilize the normalization of wavefunction
\begin{eqnarray}
\sum_{x}{|u_{0,k}(x_0)u_{0,k}(x)|^2}&=&|u_{0,k}(x_0)|^2\sum_{x}{|u_{0,k}(x)|^2} \notag \\ 
&=&|u_{0,k}(x_0)|^2. \notag 
\end{eqnarray}
By assuming the interaction strength $g$ is much weaker than the bandwidth of $E_{0,k}$, the Weisskopf-Wigner approximation is valid. The variable $|u_{0,k}(x_0)|^2$ can be seen as a constant $|u_{0,k_r}(x_0)|^2$. Then, we have
\begin{gather}
\dot{c}_e\left( t \right) =-\frac{g^2}{2\pi}\sum_{\pm}|u_{0,\pm k_r}\left( x_0 \right) |^2\int_{-\pi}^{\pi}{dk}\int_0^t{c_e\left( t' \right) e^{-i\Delta _k\left( t-t' \right)}dt'}. \label{dot_c_e}
\end{gather}
Note that $\Delta _k = E_{0,k}-w_q$. As depicted in Fig.~2(b), around $E_{0,k_r}=w_q$ the dispersion relation $E_{0,k}$ is linear with a group velocity $v_g=\frac{\partial}{\partial k}E_{0,k}$, i.e., $\Delta _k = E_{0,k}-E_{0,k_r}= v_g(k-k_r)=v_g\delta k.$ Eq.~(\ref{dot_c_e}) is rewritten as 
\begin{eqnarray}
\dot{c}_e\left( t \right)= &&-\frac{g^2}{2\pi}\sum_{\pm}|u_{0,\pm k_r}\left( x_0 \right) |^2 \notag \\
&& \int_{-\pi}^{\pi}{\frac{1}{v_g}d\left( v_g\delta k \right)}\int_0^t{c_e\left( t' \right) e^{-iv_g\delta k\left( t-t' \right)}dt'}. \label{c_e_step}
\end{eqnarray}
The integral bound $ \pm \pi$ can be extended to infinity. The corresponding solution to Eq.~(\ref{c_e_step}) is 
\begin{gather}
c_e\left( t \right) =e^{-\frac{\varGamma}{2}t},\quad \varGamma=\sum_{\pm}\varGamma _{\pm}=\sum_{\pm}\frac{g^2}{v_g}|u_{0,\pm k_r}\left( x_0 \right) |^2, \label{rate}
\end{gather}
where $\varGamma_{+}\ (\varGamma_{-})$ corresponds to the emission rate into $+y\ (-y)$ direction.

\begin{figure*} 
	\begin{center}
		\centering \includegraphics[width=16.8cm]{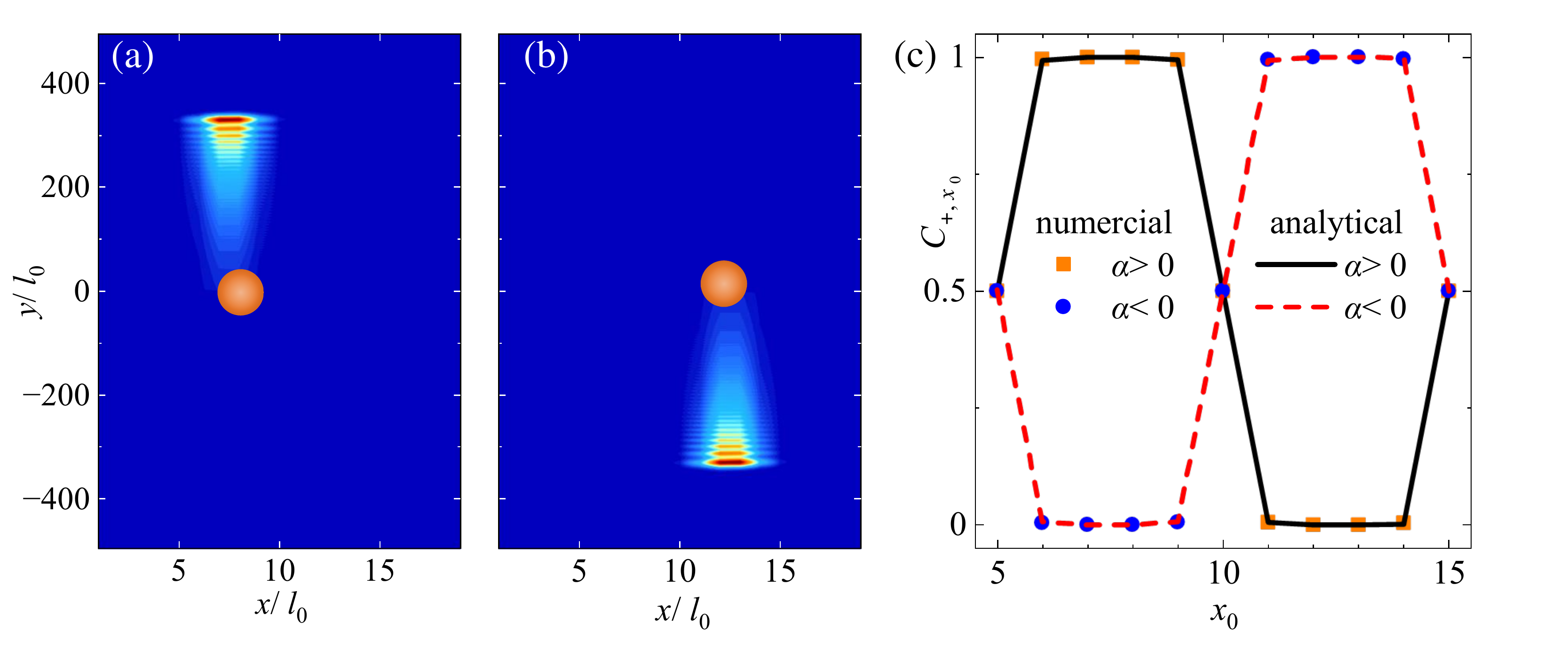}
		\caption{(a) and (b) The photonic field $|c_{x,y}|^2$, for the emitter coupling to $(x_0=8,y_0=0)$ and $(x_0=12,y_0=0)$ (the dot). (c) The chiral factors $C_{+}$ for $\alpha > 0$ and $\alpha < 0$ are calculated by numerical simulation via Eq.~(\ref{chiralfactor}) and analytical via Eq.~(\ref{Anachiralfactor}). The parameters for those plots are $\omega_q=-3.42$, $|\alpha|=0.1$, $\lambda=0.01$ and $g=0.005$.}
		\label{fig4}
	\end{center}
\end{figure*}

We simulate the interaction between a two-level emitter and a $N_x \times N_y$ square lattice in real space. The Hilbert space is restricted within the single-excitation subspace, i.e.,
\begin{gather}
|\Psi(t) \rangle =c_e\left( t \right) |e,0\rangle +\sum_{x=1}^{N_x}\sum_{y=1}^{N_y}{c_{x,y}\left( t \right)}|g,x,y\rangle. \label{basis}  
\end{gather}
Here $c_e(t)$ denotes the amplitude of the emitter in the excited state and $c_{x,y}(t)$ is the amplitude of a single photon at $(x,y)$ in the lattice. The Hamiltonian in Eq.~(\ref{H_int}) can be expanded in the basis of Eq.~(\ref{basis}). Then, by numerically solving the Schrödinger equation, we obtain the probability of the emitter $|c_{e}(t)|^2$ and the photonic field $|c_{x,y}(t)|^2$.  For a finite $N_x\times N_y$ lattice, the Hilbert space is $N_xN_y+1$. We set $N_x=25$ and $N_y=1000$, which is large enough to avoid the propagation field touching the lattice boundary.  

The spectrum of the extended H-model is dispersive, with a bandwidth proportional to $\lambda$. Therefore, when $\lambda \gg g$, the Markovian approximation is valid. In Fig.~\ref{fig3}(a), we plot the evolution of the emitter via numerical simulation, which matches well with an exponential decay $e^{-\varGamma t}$ [$\varGamma$ is given by Eq.~(\ref{rate})]. When $\lambda$ is comparable to $g$, the Markovian approximation is not valid. The modes around the band edge with zero group velocity contribute significantly to the dynamics of the emitter~\cite{Calaj2016,Zhao2020}. The dynamic evolution is a fractional decay process~\cite{Cirac2017PRA,Cirac2017PRL,Lombardo2014,Cascio2019,Garc2020}. Partial energy is trapped around the coupling position, and cannot propagate. The trapped energy oscillates between the emitter and the lattice sites around the coupling position~\cite{Shi2016,Sundaresan2019,Rom2020,Wang2021PRl,Kim2021}. For $\lambda=0$ the whole photonic energy is trapped around the coupling position, and complete Rabi oscillation is observed, as discussed in Ref.~\cite{Bernardis2021}. We plot the non-Markovian evolution of the emitter with $\lambda=0$ and $\lambda=10^{-4}$ in Fig.~\ref{fig3}(b).
 
When the Markovian approximation is valid, we plot in Fig.~\ref{fig3}(c) the decay rate versus the coupling position $x_0$ via numerical simulation. The analytical results [Eq.~(\ref{rate})] are also plotted. The rate has a periodic relationship with $x_0$, which stems from the wavefunction of the coupling position, $|u_{0,k_r}(x_0)|^2$. For example, we set $k_r=\pi/2$ and plot the probability $|u_{0,\pm k_y}(x_0)|^2$ for different $x_0$ in Fig.~\ref{fig3}(d). Note that the $\sum_{\pm} |u_{0,\pm k_y}(x_0)|^2$ is symmetric about $x_0=10$. According to Eq.~(\ref{rate}), the decay rate of the emitter is proportional to $\sum_{\pm} |u_{0,\pm k_y}(x_0)|^2$. Eventually, the decay rate changes with the coupling position periodically. The lengthscale of the oscillation is $1/\alpha$. 

Then we pay more attention to the difference between $\varGamma_{\pm}$. From Eq.~(\ref{rate}) and Fig.~\ref{fig3}(d), $\varGamma_{+} \ne \varGamma_{-}$ except $x_0=5$ and $x_0=10$, which leads to unidirectional emission along y-axis. In the following, we will focus on the chiral emission along y-axis.

\section{CHIRAL FIELD PROPAGATION}
The decay rates $\varGamma_{\pm}$ are not identical. For an emitter located at the position $(x_0,0)$, the positive chiral factor is defined as
\begin{gather}
\mathcal{C}_{+, x_0}=\frac{\Phi _+}{\sum \Phi _{\pm}}=\frac{\varGamma_{+k_r} ( x_0 )}{\sum{\varGamma_{\pm k_r} ( x_0 )}}, \label{chiralfactor}
\\ 
\Phi _{\pm}=\sum_{x}{\sum_{y=0}^{\pm N_y/2}{|c_{x,y}|^2}}.
\end{gather}
According to Eq.~(\ref{rate}) and Eq.~(\ref{chiralfactor}), the analytical chiral factor is derived as
\begin{gather}
\mathcal{C} _{+, x_0}=\frac{|u _{0,+k_r}\left( x_0 \right) |^2}{\sum_{\pm}{|u_{0,\pm k_r}\left( x_0 \right) |^2}}.
\label{Anachiralfactor} 
\end{gather}
\begin{figure} 
	\begin{center}
		\centering \includegraphics[width=8.5cm]{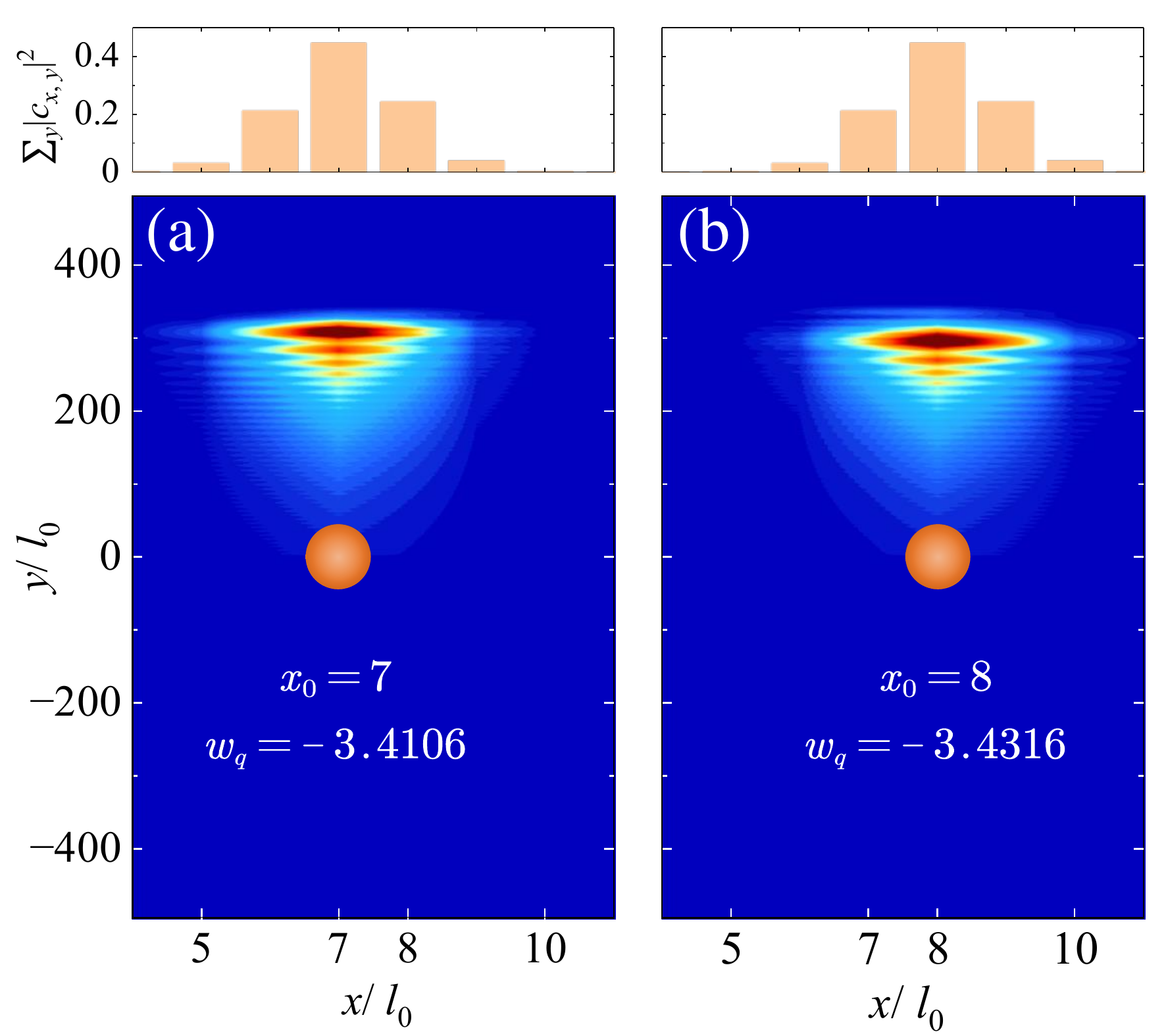}
		\caption{The photonic field $|c_{x,y}|^2$. (a) the emitter is located at $x_0=7$ (the dot), with the frequency $-3.4106$. (b) the emitter is located at $x_0=8$, with $w_q=-3.4316$. The parameters for those plots are $\alpha=0.1$, $\lambda=0.01$ and $g=0.005$.}
		\label{fig5}
	\end{center}
\end{figure}
In Fig.~\ref{fig4}(a, b), we plot the photonic field for the emitter located at $(x_0=8,\ y_0=0)$ and $(x_0=12,\ y_0=0)$, respectively. Note that the field propagates unidirectionally. Moreover, the chiral factor depends on the coupling position $x_0$. In Fig.~\ref{fig4}(c), we show how $\mathcal {C}_{+,x_0}$ changes with $x_0$. The length of the period is equal to $1/\alpha = M$. For some points, the chiral factor can reach $1$.

More intriguingly, the direction of the chiral bulk states depends on the signs of $\alpha$ and $\lambda$. We plot the chiral factor for different $\alpha$ in Fig.~\ref{fig4}(c). By switching the sign of the parameter $\alpha$, the chiral emission of the bulk state is reversed. It's because that the $|u_{0,x=M-m}^{\alpha>0}(k_y)|^2$ is transformed into $|u_{0,x=m}^{\alpha<0}(k_y)|^2$, and $\varGamma^{\alpha>0}_{\pm}=\varGamma^{\alpha>0}_{\mp}$. Furthermore, changing the signs of $\lambda$ leads to the dispersion relation with opposite group velocities $v_g(\pm k_r)$, as shown in Fig.~\ref{fig2}(b). Additionally, the sign of group velocity $v_g(\pm k_r)$ denotes the direction of wave pocket propagation. The chiral emission direction is reversed by flipping $\lambda$. Therefore, we can modulate the sign of $\lambda$ and $\alpha$ to change emission direction along $y$-axis.

Note that when the emitter is resonant with the middle of $E_{0,k}$, i.e., $k_r=\pi/2$, we have $|u_{0,k_r}(7)|^2 \simeq |u_{0,k_r}(8)|^2$. No matter the emitter is located at $x=7$ or $x=8$, the photon current mainly propagates along two lines $x=7$ and $x=8$, as shown in Fig.~\ref{fig4}(a). According to numerical results, when $|u_{0,+k_r}(x=7)|^2$ reaches its maximum, the emitter's frequency should be set as $w_q=-3.4106$. In that case, we set the emitter located at $x_0=7$ and most of the photonic field distributes on $x=7$, as shown in Fig.~\ref{fig5}(a). In Fig.~\ref{fig5}(b), when $w_q=-3.4316$ and $x_0=8$, most of the field distributes on $x=8$. Therefore, we can control the photonic field's distribution by adjusting the emitter's frequency.

\section{conclusion}

In this work, we have explored the dynamic evolution of an emitter coupling to an extended Hofstadter model with the NNN couplings. The NNN hopping breaks the mirror symmetry, leading to non-flat bands with nonzero group velocities. By considering the emitter located in the bulk region of the extended H-model, we observe a chiral Markovian decay process without coherent oscillations, which is totally different from the phenomena in Ref.~\cite{Bernardis2021}. The physical mechanism of the chiral emission arises from the asymmetry of the wave function in $k$ space. Moreover, the decay rate and the chiral factor change with the coupling position periodically. The period is the reciprocal of the effective magnetic flux. The chiral factor can reach $1$ by choosing the proper coupling position. Moreover, the directional decay can be controlled by the NNN couplings.

In the artificial quantum systems, the NNN couplings exist widely. For example, in bilayer or doped graphene, the NNN hopping strength is around $5\%$ of the NN interaction strength~\cite{Wallace1947,Reich2002}. Hence, our work will provide guidance for the experimental realization of exotic quantum dynamics in the extended H-model. We believe our proposal can be also utilized to simulate the quasi-1D chiral photon transport in 2D system.

\section{Acknowledgments}
The quantum dynamical simulations are based on open source code 
QuTiP~\cite{Johansson12qutip,Johansson13qutip}. 
X.W.~is supported by 
the National Natural Science
Foundation of China (NSFC; No.~12174303 and Grant No.~11804270), and the 
Fundamental 
Research Funds for the Central Universities (No. xzy012023053). WXL is 
supported by the Natural Science Foundation of Henan
Province (No. 222300420233).

\appendix
\section{MIRROR SYMMETRY}
In quantum mechanics, if the Hamiltonian and the operator satisfy the commutation relation, i.e., $[H,\ A]=0$, the system has the corresponding symmetry. Regarding mirror symmetries, the operator $M_x\ (M_y)$ seeds $x\ (y) \rightarrow -x\ (-y)$ and the Hamiltonian obeys $[H,\ M_{x(y)}]=0$. However, in Hofstadter model, due to the magnetic flux $\phi =2\pi\alpha $, the operator $M_{x(y)}$ must multiply a gauge transformation $\mathcal{G}$ to recover the symmetry, $M_{x(y)}^{\alpha}=M_{x(y)}^{\alpha=0}\mathcal{G}^{\alpha}$~[76]. In this scenario, the Hamiltonian does not obey the original relation for mirror symmetry, but the relation with flux reversal, i.e.,
\begin{gather}
(M_i^{\alpha})H^{\alpha}(M_{i}^{\alpha})^{\dagger}=H^{-\alpha}. \label{M_Mirror_symm}
\end{gather}
When we consider the NNN couplings, Eq.~(\ref{M_Mirror_symm}) is not valid. The mirror symmetries are broken. To explain this point clearly, we take the model in the main text for example.

For the whole square lattice ($N\times N$), the Hamiltonian is 
\begin{widetext}
	\begin{gather}
	H_s=\left( \begin{matrix}
	H_x&		H_y&		0&		\cdots&		H_{y}^{\dagger}\\
	H_{y}^{\dagger}&		H_x&		H_y&		\cdots&		0\\
	0&		H_{y}^{\dagger}&		H_x&		\cdots&		0\\
	\vdots&		\vdots&		\vdots&		\ddots&		\vdots\\
	H_y&		0&		0&		H_{y}^{\dagger}&		H_x\\
	\end{matrix} \right),\quad H_x=\left( \begin{matrix}
	0&		1&		0&		\cdots&		1\\
	1&		0&		1&		\cdots&		0\\
	0&		1&		0&		\cdots&		0\\
	\vdots&		\vdots&		\vdots&		\ddots&		\vdots\\
	1&		0&		0&		1&		0\\
	\end{matrix} \right) ,\quad H_y=\left( \begin{matrix}
	e^{i\phi _1}&		\lambda&		0&		\cdots&		\lambda\\
	\lambda&		e^{i\phi _2}&		\lambda&		\cdots&		0\\
	0&		\lambda&		e^{i\phi _3}&		\cdots&		0\\
	\vdots&		\vdots&		\vdots&		\ddots&		\vdots\\
	\lambda&		0&		0&		\lambda&		e^{i\phi _N}\\
	\end{matrix} \right) , \label{H_matrix}
	\end{gather}
\end{widetext}
where $H_x\ (H_y)$ is a $N\times N$ square matrix, and $H_s$ is a $N^2 \times N^2$ square matrix. The sub-diagonal terms of $H_x$ are the NN  hopping along $x$ direction. The diagonal terms of $H_y$ are the NN hopping along $y$-direction with the hopping phase $e^{i\phi_{n}}$ with $\phi_{n}=2\pi\alpha n$. The sub-diagonal terms of $H_y$ are the NNN couplings both in the $x$ and $y$ directions. The elements in the top right and bottom left corner of $H_s$, $H_x$ and $H_y$, are the periodic boundary conditions. 

We define matrices
\begin{gather}
M_1=\left( \begin{matrix}
1&		0&		\cdots&		0\\
0&		1&		\cdots&		0\\
\vdots&		\vdots&		\ddots&		\vdots\\
0&		0&		0&		1\\
\end{matrix} \right) ,\quad M_2=\left( \begin{matrix}
0&		0&		0&		1\\
0&		0&		1&		0\\
\vdots&		\ddots&		\vdots&		\vdots\\
1&		0&		0&		0\\
\end{matrix} \right) ,
\\
M_3=\left( \begin{matrix}
e_1&		0&		\cdots&		0\\
0&		e_2&		\cdots&		0\\
\vdots&		\vdots&		\ddots&		\vdots\\
0&		0&		0&		e_N\\
\end{matrix} \right) ,
\end{gather}
where $e_n=e^{i2\pi(1-\alpha n)}$. The mirror symmetry matrices, for $\alpha = 0$, are $M_x^0=M_1 \otimes M_2$ and $M_y^0=M_2 \otimes M_1$. Under the Landau gauge, the gauge transform is $\mathcal{G}=M_1 \otimes M_3$. Then $M_x^{\alpha}=M_x^0\mathcal{G}$, and $M_y^{\alpha}=M_y^0$. Therefore, Hamiltonian $H_s(\alpha,\lambda)$ obeys
\begin{gather}
[M_{x(y)}^0]^{\dagger}H_s(0,\lambda)[M_{x(y)}^0]=H_s(0,0), 
\\
[M_{x(y)}^{\alpha}]^{\dagger}H_s(\alpha,0)[M_{x(y)}^{\alpha}]=H_s(-\alpha,0),
\\
[M_{x}^{\alpha}]^{\dagger}H_s(\alpha,\lambda)[M_{x}^{\alpha}] \ne H_s(-\alpha,\lambda), \label{mirror_broken}
\\
[M_{y}^{\alpha}]^{\dagger}H_s(\alpha,\lambda)[M_{y}^{\alpha}]=H_s(-\alpha,\lambda).
\end{gather}
Equation~(\ref{mirror_broken}) indicates that the NNN couplings break the mirror symmetry, with $\alpha \ne 0$, which is the mechanism for the chiral emission in the main text.

\section{QUASI-CONTINUES HARPER-EQUATION}
In order to obtain the analytical dispersion relation of the extended H-model, we adopt the quasi-continuous approximation $u(x\pm1)\rightarrow e^{\pm \partial_{x}}u(x)$ in Eq.~(\ref{Harper_equation})~\cite{Harper2014}. We assume that the hopping is related to the Laplacian operator via a finite difference approximation. The eigen-equation is rewritten as
\begin{eqnarray}
E u\left( x \right)=&-& \big[(1+2\lambda \cos k_y)( \hat{T}_-+\hat{T}_+ ) \notag \\
&+&  2\cos \left( 2\pi \alpha x-k_y \right) \big]  u \left( x \right), 
\label{quasicontinuewavefunction}
\end{eqnarray}
with $\hat{T}_{\pm}=e^{\pm \partial _x}$. When $x'=x+\frac{1}{\alpha}=x+M$, the coefficients of $u(x')$ are totally equal to $u(x)$, indicating that this system has periodicity with $M$ as a period. For cosine term, we adopt $x=x'+k_y/2\pi \alpha$ to remove $k_y$, i.e., $\cos (2\pi \alpha x-k_y)=\cos(2\pi \alpha x')$. Using the Taylor's expansion, 
\begin{gather}
e^{\pm \partial _x}=1+\frac{\pm \partial _x}{1!}+\frac{\left( \pm \partial _x \right) ^2}{2!}+\mathcal{O} \left( x^3 \right), \notag \\
\cos \left( x \right) =1-\frac{x^2}{2!}+\mathcal{O} \left( x^4 \right), \notag
\end{gather}
the equation is written as:
\begin{gather}
\frac{E+2+2\lambda_0}{2 \sqrt{\lambda _0} }=-\frac{1}{2}\sqrt{\lambda _0} \partial _{x}^{2}+\frac{1}{2}\frac{\left( 2\pi \alpha \right) ^2}{\sqrt{\lambda _0}}x^2,\label{oscillator}
\end{gather}
which $\lambda_0=1+2\lambda\cos k_y$. We define $w=2\pi \alpha $, $m=\frac{1}{\sqrt{\lambda _0}}$, $\hat{p}=-i\partial _x$, $\hat{x}=x$ and $a^{\dagger}=\sqrt{\frac{mw}{2}}\hat{x}-\frac{i}{\sqrt{2mw}}\hat{p}
$. The Eq.~(\ref{oscillator}) is similar with the equation of a harmonic oscillator. By considering the higher order terms, we obtain the dispersion relations in Eq.~(\ref{dispersing_relation1}).

\section{DISCRETE WAVE FUNCTION}
Although the dispersion relations are derived through the quasi-continuous approximation, the discrete wavefunction cannot be obtained. Hence, via Fourier transformation in y-axis $$a_{x,y}^{\dagger}=\frac{1}{\sqrt{N}}\sum_{k}{e^{-ik_y y} a_{x,k_y}^{\dagger}},$$ we rewrite $H_m$ in $k_y$ space as 
\begin{gather}
H_m(k_y)=-\sum_x \left( 1+2\lambda \cos k_y \right) \left( a_{x+1,k_y}^{\dagger}a_{x,k_y} \right. \notag \\      \left. +a_{x-1,k_y}^{\dagger}a_{x,k_y} \right) - 2\cos \left( 2\pi \alpha x+k_y \right) a_{x,k_y}^{\dagger}a_{x,k_y} +\mathrm{H}.\mathrm{c}.
\end{gather}
Under the periodic boundary conditions and adopting $\alpha=1/M$, the Hamiltonian $H_m(k_y)$ can be decomposed as a $M$-dimensional Hermitian matrix
\begin{gather}
H_m(k_y)=-\left[ \begin{matrix}
h_1&		h_c&		0&		\cdots&		h_c\\
h_c&		h_2&		h_c&		\cdots&		0\\
0&		h_c&		h_3&		\cdots&		0\\
\vdots&		\vdots&		\vdots&		\ddots&		\vdots\\
h_c&		0&		0&		\cdots&		h_M\\
\end{matrix} \right],
\end{gather} 
where $h_i=2\cos \left( 2\pi \alpha i + k_y \right)$ and $h_c=1+2\lambda \cos  k_y$.
Using $U^{-1}H_m(k_y)U=E$, we exactly diagonalize the Hamiltonian. The eigenvalue $E_l$ is the $l$th diagonal elements of $E$ and the corresponding wave function $u_{l,k_y}$ is the $l$th column vector of $U$, where $l$ denotes the $l$th energy level. The wave functions and $U$ matrix are expressed as
\begin{gather}
u_{l,k_y}=[u_{l,k_y}(1),u_{l,k_y}(2),...,u_{l,k_y}(M)]^T ,\notag 
\\
U=[u_{1,k_y},u_{2,k_y},...,u_{M,k_y}].
\end{gather}
Consequently, the eigenmodes are
\begin{gather}
X_{l,k_y}=\left[ u_{l,k_y}(1)a_{1,k_y},  u_{l,k_y}(2)a_{2,k_y},..., u_{l,k_y}(m)a_{M,k_y}\right].
\end{gather}
The eigenmodes and the annihilation(creation) operators satisfy the following relationships
\begin{gather}
\left[ X_{1,k_y}^{\dagger}, X_{2,k_y}^{\dagger},... \right] =\left[ a_{1,k_y}^{\dagger}, 
a_{2,k_y}^{\dagger},... \right] \times U, \hfill
\notag \\
\left[ a_{1,k_y}^{\dagger}, a_{2,k_y}^{\dagger},... \right]=\left[ X_{1,k_y}^{\dagger}, 
X_{2,k_y}^{\dagger},... \right] \times U^{-1}\hfill .
\end{gather}
Since the Hamiltonian $H_m(k_y)$ is Hermitian, $U$ is a unitary matrix i.e., $U^{-1}=U^{T}$ and $$\left[ u_{1,k_y}, u_{2,k_y}, ... \right]^{-1}=\left[ u_{1,k_y}, u_{2,k_y}, ... \right]^{T}.$$ Eventually, the annihilation operators in real space can be rewritten as
\begin{eqnarray}
a_{x,y}^{\dagger}&=&\frac{1}{\sqrt{N}}\sum_{k}{e^{-ik_y y}a_{x,k_y}^{\dagger}} \notag \\
&=&\frac{1}{\sqrt{N}}\sum_{k}\sum_{i=1}^M{e^{-iky} {u_{i,k}\left( x \right) 
X_{i,k}^{\dagger}}}\label{realmodeleigenmode}.
\end{eqnarray}


%

\end{document}